\documentclass[referee]{aa}
\usepackage{graphics}
\usepackage{epsfig}
\usepackage{txfonts}

\newcommand{\lab}{\label}
\begin{document}
\title{Accelerating universe in scalar tensor models -
confrontation of theoretical predictions with observations.}
\author{M.\,Demianski \inst{1} \fnmsep \inst{2}, E.\,Piedipalumbo\inst{3}\fnmsep \inst{4}, C.\,Rubano\inst{3}\fnmsep \inst{4}
, C. Tortora \inst{3}\fnmsep \inst{4} } \offprints{E.Piedipalumbo,
ester@na.infn.it} \institute{Institute for Theoretical Physics,
University of Warsaw,  Hoza 69, 00-681 Warsaw, Poland \and
Department of Astronomy, Williams College, Williamstown, Ma 01267,
USA \and Dipartimento di Scienze Fisiche, Universit\`{a} di Napoli
Federico II, Compl. Univ. Monte S. Angelo, 80126 Naples, Italy
\and Istituto Nazionale di Fisica Nucleare, Sez. Napoli, Via
Cinthia, Compl. Univ. Monte S. Angelo, 80126 Naples, Italy }
\titlerunning{Accelerating universe in scalar tensor models...}
\authorrunning{M.\,Demianski \& al.}
\date{Received / Accepted}
\abstract{}{To study the possibility of appearance of accelerated
universe in scalar tensor cosmological models}{We consider scalar
tensor theories of gravity assuming that the scalar field is non
minimally coupled with gravity. We use this theory to study
evolution of a flat homogeneous and isotropic universe. In this case
the dynamical equations can be derived form a point like Lagrangian.
We study the general properties of dynamics of this system and show
that for a wide range of initial conditions such models lead in a
natural way to an accelerated phase of expansion of the universe.
Assuming that the point like Lagrangian admits a Noether symmetry we
are able to explicitly solve the dynamical equations. We study one
particular model and show that its predictions are compatible with
observational data, namely the publicly available data on type Ia
supernovae, the parameters of large scale structure determined by
the 2-degree Field Galaxy Redshift Survey (2dFGRS), the measurements
of cosmological distances with the Sunyaev-Zel'dovich effect and the
rate of growth of density perturbations}{ It turns out that this
model have a very interesting feature of producing in a natural way
an epoch of accelerated expansion.  With an appropriate choice of
parameters our model is fully compatible with several observed
characteristics of the universe}{}\maketitle

\section{Introduction}
Recent observations of the type Ia supernovae, Gamma Ray Bursts,
and CMB anisotropy indicate that the total matter-energy density
of the universe is now dominated by some kind of dark energy
(\cite{rie+al98,Riess00,Riess04}). The origin and nature of this
dark energy is not yet known~(\cite{zel67,weinberg2}).

In the last several years a new class of cosmological models has
been proposed. In these models the standard cosmological constant
$\Lambda$-term is replaced by a dynamical, time-dependent component
- quintessence or dark energy - that is added to baryons, cold dark
matter (CDM), photons and neutrinos. The equation of state of the
dark energy is given by $w_{\phi} \equiv p_{\phi}/ \rho_{\phi}$,
where $p_{\phi}$ and $\rho_{\phi}$
 are, respectively, the pressure and energy density,
and $-1 \leq w_{\phi} <0$, what implies a negative contribution
to the total pressure of the cosmic fluid. When $w_{\phi} =-1$,
we recover a constant $\Lambda$-term. One of the possible
physical realizations of quintessence is a cosmic scalar field
(\cite{cal+al98}), which induces dynamically a repulsive
gravitational force, causing  an accelerated expansion of the
Universe, as recently discovered by observations of distant type
Ia supernovae (SNIa) (\cite{rie+al98,Riess04}) and confirmed by
the WMAP observations (\cite{spergel}).

The existence of a considerable amount of dark energy leads to at
least two theoretical problems: 1) why only recently dark energy
started to dominate over matter, and 2) why during the radiation
epoch the density of dark energy is vanishingly small in
comparison with the energy density of radiation and matter ({\it
fine tuning} problem). The fine tuning problem can be alleviated
by considering models of dark energy that admit so called {\it
tracking behavior} (\cite{stein2}). In such models, for a wide
class of initial conditions, equation of state of dark energy
tracks the equation of state of the background matter and
radiation (\cite{stein2,zlat}). All these circumstances stimulated
a renewed interest in the generalized gravity theories, and
prompted consideration of a variable $\Lambda $ term in more
general classes of theories, such as the scalar tensor theories of
gravity (\cite{francesca1}).  One of the additional advantages of
appealing to these theories is that they open new prospective in
the scenario of a {\it decaying dark energy}, since the same field
that is causing the time (and space) variation of the dark energy
is also causing the Newton's constant to vary.

In this paper we consider cosmological models in a non minimally
coupled scalar tensor theory of gravitation.  We assume that the
universe is homogeneous and isotropic and its geometry is described
by the Friedman-Robertson-Walker line element. For the reason of
simplicity we consider only the case of flat universe. In this case
the scalar field depends only on time and the dynamical equations
that describe evolution of the geometry and the scalar field can be
derived form a point like Lagrangian. We derive the general set of
dynamical equations and discuss their basic properties. When we
require that the point like Lagrangian admits an additional Noether
symmetry the dynamical equations can be explicitly integrated. In
particular we consider a model with the scalar field potential of
the form $V(\phi)=V_0\phi^4$ and we analyze its dynamics. Finally to
compare predictions of our model with observations we concentrate on
the following data: the publicly available data on type Ia
supernovae, the parameters of large scale structure determined by
the 2-degree Field Galaxy Redshift Survey (2dFGRS), and measurements
of cosmological distances with the Sunyaev-Zel'dovich effect. We
show that our model is compatible with these observational data.

Our paper is organized as follows: in Sec. 2 we present our model
and discuss its basic properties. In Sec. 3 we confront predictions
of our model with observational data, Sec. 4 is devoted to the
discussion of evolution of density perturbations in our model and
finally in Sec. 5 we present our conclusions.

\section{Model description}
Let us consider the general action of a scalar field $\phi$ non
minimally coupled with gravity when there is no coupling between
matter and $\phi$
\begin{equation}
{\cal A} = \int_T \sqrt{-g} \left( F(\phi) R +
\frac{1}{2}g^{\mu\nu} \phi_{, \mu} \phi_{,\nu} - V(\phi) + {\cal
L}_m \right)d^{4}x \,, \label{e1}
\end{equation}
where $F(\phi),~ V(\phi)$ are two generic functions representing
the coupling of the scalar field with geometry and its potential
energy density respectively, $R$ is the curvature scalar,
${\displaystyle \frac{1}{2}g^{\mu\nu} \phi_{, \mu} \phi_{, \nu}}$
is the kinetic energy of the scalar field $\phi$ and ${\cal L}_m$
describes the standard matter content. In units such that $8 \pi
G_{N}=\hbar= c = 1$, where $G_{N}$ is the Newtonian constant, we
recover the standard gravity when $F=-\displaystyle{1\over 2}$,
while in general the effective gravitational coupling
$G_{eff}=-{1\over {2F}}$. Here we would like to study the simple
case of a homogeneous and isotropic universe, what implies that
the scalar field $\phi$ depends only on time. It turns out that in
the flat Friedman-Robertson-Walker cosmologies, the action in Eq.
(\ref{e1}) reduces to the {\em pointlike} Lagrangian\footnote{We
use the expression {\em pointlike} to stress that the field
Lagrangian obtained from Eq. (\ref{e1}) can be considered as
defined in the minisuperspace where the remaining two variables
($a, \phi$) depend only on the cosmological time $t$ and so they
can be considered as describing a mechanical system with
two-degrees of freedom.}
\begin{equation}
{\cal L}= 6 F a \dot{a}^2 + 6 F' \dot{\phi}a^2 \dot{a}+
a^3\left({1\over 2} \dot{\phi}^2-V(\phi)\right) - D a^{-3(\gamma
-1)} \,, \label{e2}
\end{equation}
where $a$ is the scale factor and  prime denotes derivative with
respect to $\phi$, while dot denotes derivative with respect to
time. Moreover, the constant D is defined in such a way that the
matter density $\rho_{m}$ is expressed as $\rho_m= D (a_o/
a)^{3\gamma}$, where $1 \leq \gamma \leq 2$. The effective
pressure and energy density of the $\phi$-field are given by
\begin{equation}
p_{\phi}= \frac {1}{2} \dot{\phi}^2- V(\phi)- 2(\ddot{F}+ 2H
\dot{F}) \,, \label{fi-pressure}
\end{equation}
\begin{equation}
\rho_{\phi}= \frac {1}{2} \dot{\phi}^2+ V(\phi)+ 6 H \dot{F} \,,
\label{fi-density}
\end{equation}
where  $H=\displaystyle{{\dot a}\over a}$  is the Hubble constant.
These two expressions, even if not pertaining to a conserved
energy-momentum tensor, define an effective equation of state
$w_\phi=\displaystyle {p_{\phi}\over \rho_{\phi}}$, which drives
the late time behavior of the model. The field equations derived
from Eq. (\ref{e2}) are the same ones which would come from the
field equations derived from Eq. (\ref{e1}) when homogeneity and
isotropy is imposed, that is
\begin{equation}
H^2= -\frac{1}{2 F} \left( \frac{\rho_\phi}{3}+
\frac{\rho_m}{3}\right) \,, \label{e3}
\end{equation}
\begin{equation}
2 \dot{H}+ 3 H^2= \frac{1}{2 F} (p_{\phi}+ p_m) \,, \label{e4}
\end{equation}
and
\begin{equation}
\ddot{\phi}+ 3 H \dot{\phi}+ 6( \dot{H} + 2H^{2})F'+ V' =0\,.
\label{klein-gordon}
\end{equation}
The scalar tensor theory with the scalar field non minimally
coupled to gravity provides a wide framework to study
cosmological models. The function $F(\phi)$ that describes the
coupling between the scalar filed and gravity is influencing not
only the evolution of the cosmological scale factor and the
scalar field itself but also determines the strength of
gravitational interactions. In this paper we consider only the
simple case of a homogeneous and isotropic flat universe filled
in with a scalar field (quintessence) and pressureless matter,
i.e., $p_m =0$ (dust). That is, strictly speaking, our model
describes the evolution of the universe only after the
matter-radiation decoupling. The scalar field which appears in
our model is treated as quintessence. To discuss the changing
influence of the scalar field on the evolution of the universe
and the effective equation of state of quintessence, we set
$x\equiv\displaystyle{{\dot\phi}^2\over 2V}$, and $\displaystyle
x_1\equiv {\dot{F}\over V}$, we find that
\begin{eqnarray}\label{wphi}
  w_\phi={p_{\phi}\over \rho_{\phi}} &=&{ x-1+2x_1\left({\dot{V}\over V}+2H\right) -2\dot{x}_1\over x+1 +6H x_1}\,.
\end{eqnarray}
The parameter $x$ measures the ratio of the kinetic energy relative to the potential energy of the scalar field. In the non minimally coupled
case it is possible to invert the Eq.(\ref{wphi}) and we get that
\begin{equation}\label{xb}
    x={\rho_\phi(1 + w_\phi)+2\left(\ddot{F}-H \dot{F}\right)\over \rho_\phi(1 - w_\phi)-2 \left(\ddot{F}+ 5H \dot{F}\right)}\,.
\end{equation}
When the coupling function $F$ is a constant we recover the
relation $x\displaystyle={1+w_\phi\over 1-w_\phi}$ that holds in
the minimally coupled case.
When $w_\phi<0$ the quintessence contributes negative pressure,
this occurs when
\begin{equation}\label{wneg}
x<1+2\left[\dot{x}_1-x_1\left({\dot{V}\over V}+2 H\right)\right],
\end{equation}
while the inequalities
\begin{eqnarray}
  x & \geq & \dot{x}_1-x_1\left({\dot{V}\over V}+5H\right)\label{nquint}\,, \\
   x & < & \dot{x}_1-x_1\left({\dot{V}\over
   V}+5H\right)\label{supquint}\,,
\end{eqnarray}
correspond to the {\it standard} quintessence $(w_\phi\geq -1)$
and superquintessence respectively $(w_\phi<-1)$. If both
$F(\phi)$ and $V(\phi)$ are known, it is possible to describe, in
the parameter space, the transition between standard quintessence
and superquintessence. Let us now introduce the concept of an
effective cosmological constant $\Lambda_{eff}$. Using
Eq.(\ref{e3}) it is natural to define the effective cosmological
constant as $\Lambda_{eff}=-\displaystyle{\rho_{\phi}\over {2F}}$.
With this definition we can rewrite Eq.(\ref{e3}) as
\begin{eqnarray}
  3H^2 &=& G_{eff}\rho_m +\Lambda_{eff}\,.
\end{eqnarray}
Introducing the standard Omega parameters by
\[ \Omega_{m}=-{\rho_{m}\over {6FH^{2}}}, \quad
\Omega_{\phi}={\Lambda_{eff}\over {3H^{2}}}=-{\rho_{\phi}\over
{6FH^{2}}} \,,\] we get that as usual
\begin{eqnarray}
   \Omega_m+\Omega_{\phi}= 1\,.
\end{eqnarray}
>From the definition of $\rho_{\phi}$ and $p_{\phi}$ and the
generalized Klein-Gordon equation it follows that
\begin{equation}\label{einstnew2}
   \dot{\rho}_{\phi}+3H\left(p_{\phi}+\rho_{\phi}\right)=-6H^2\dot{F},
\end{equation}
and
\begin{equation}\label{einstnew3}
   \dot{\Lambda}_{eff}+\dot{G}_{eff}\rho_m=-3HG_{eff}\left(p_{\phi}+\rho_{\phi}\right).
\end{equation}
These equations play a role of the continuity equations for
$\rho_{\phi}$ and $\Lambda_{eff}$.\\
We will show later on that asymptotically for large time
$\dot{G}_{eff}$ tends to zero, then $w_\phi$ determines the late
time scaling of $\Lambda_{eff}$, and actually we have that
\begin{equation}\label{asymp1}
 {\dot{\Lambda}_{eff}\over
 \Lambda_{eff}}\approx-3\tilde{H}\left(1+w_{\phi}\right),
\end{equation}
\begin{figure}
\centering{
        \includegraphics[width=9 cm, height=7 cm]{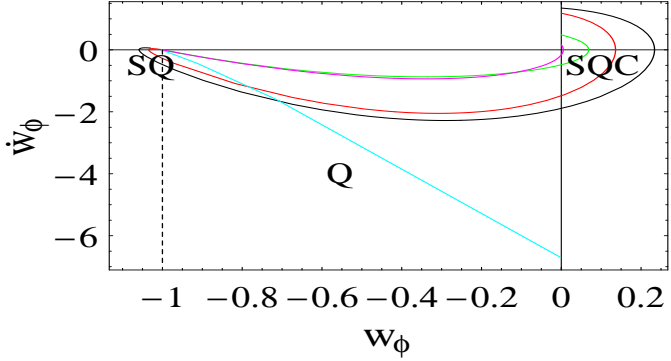}}
        \caption{Rate of change of the equation of state as measured by ${\dot w}_{\phi}$ versus the
         $w_{\phi}$ parameter. We see that together with the {\it superquintessence region}
         $w_\phi<-1$,
        (SQ), there appears also a  {\it superquintessence connected}
        region (SQC) with an equation of state $w_\phi>0$.
        The blue straight line corresponds, indeed, to values of parameters such that we
        observe neither a superquintessence expansion, nor a stiff matter behaviour.}
        \label{wacc}
\end{figure}\noindent
where $\tilde{H}$ is the asymptotic value of $H$. It turns out
that if at large $t$, $\tilde{H}$ is constant and $\neq 0$,
than asymptotically $w_\phi \rightarrow -1$. Let us note that
such a {\it transition} to the asymptotical value is responsible
for the accelerated expansion. Even if we start from a dust or
other stiff equation of state $w_\phi$ {\it converges} towards $
-1$, as it is shown in Fig.\ref{wacc}. Actually it turns out
that together with the {\it superquintessence region}
$w_\phi<-1$, (SQ), there exists also a {\it superquintessence
connected } region (SQC) with an equation of state $w_\phi>0$.
The blue straight line corresponds, indeed, to the values of
parameters such that we observe neither a superquintessence
expansion, nor a stiff matter behaviour.
\subsection{Exact solutions through Noether theorem}
To find exact solutions in the framework of the non minimally
coupled models we assume that  the {\em pointlike} Lagrangian
possesses a Noether symmetry (for a detailed exposition of this
technique which we closely follow here and interesting examples
see (\cite{rug2})). This means that we require the existence of
a Noether vector field along which the Lie derivative of the
Lagrangian is zero. This requirement restricts the possible
coupling and the form of the potential. In fact the additional
Noether symmetry exists when
\begin{equation}
V = V_0 (F(\phi))^{p(s)} \,, \label{e5}
\end{equation}
where $V_0$ is a constant and
\begin{equation}\label{ps}
p(s)= \frac{3 (s+1)}{2s +3}\,,
\end{equation}
where $s$ is a real number, and  when the coupling $F(\phi)$
satisfies the following differential equation
\begin{equation} \label{eqcoup}
d_{1}F''F^{2}+d_{2}F'^{4}+d_{3}F'^{2}F+d_{4}F^{2}=0,
\end{equation}
and the coefficients are functions of the parameter $s$,
\begin{eqnarray}&&
d_1={2s+3\over 2}\,,\\
 &&d_{2} =
3s(s+1)(s+2)\,,\\
&& d_{3} = -\frac{1}{4}(s+1)(8s^{2}+16s+3)\,,\\
&& d_{4} =\frac{s(2s+3)^{2}}{12}\,. \end{eqnarray}
A particular solution which indeed turns out to be quite
interesting is of the form
\begin{equation}
F = \xi (s) (\phi+\phi_{0})^2 \,, \label{e6}
\end{equation}
where \begin{equation} {\displaystyle \xi(s)= \frac{(2 s
+3)^2}{48 (s+1) (s+2)}}\,,\label{xi1}\end{equation} and
$\phi_{0}$ is a constant. The parameter $s$ labels then the
class of Lagrangians which admit a Noether symmetry. Let us note
that the form of the coupling given by (\ref{e6}) is quite
relevant from the point of view of fundamental physics. Not all
the values of $s$ are allowed however. From the expressions
(\ref{e5}) and (\ref{xi1}) it follows that the cases when $s=
-1$, $s= -3/2$, $s= -2$ are special and they should be treated
independently. Through Eq. (\ref{e5}) it is possible to recover
the inverse power-law potentials, while the case $s= 0$ as we
will see later is special and it corresponds to the square
hyperbolic sine potential. Both these potentials are usually
assumed {\em ad hoc} to obtain certain asymptotic behavior of
the energy density of the quintessence field (\cite{peebles,
urena}), while here they emerge naturally from the imposed
Noether symmetry. Once the general solution of Eq.(\ref{eqcoup})
is found, what automatically specifies the form of the
potential, and a generic value of $s$ is considered, it is
possible to explicitly find the Noether symmetry and to
introduce new dynamical variables associated with this symmetry
(for details see (\cite{rug2})). Using the new dynamical
variables it is then possible to solve the corresponding
Lagrange equations and finally by inverting them we obtain the
sought after $a(t)$ and $\phi(t)$. The final result can be
written in the form
\begin{equation}
 \label{general}
 a(t)=A(s)\left(B(s){(s+3)^{2}\over {6(s+6)}}t^{3\over
 {s+3}}+{D\over {\Sigma_{0}}}\right)^{{s+1}\over
s}t^{{2s^{2}+6s+3}\over {s(s+3)}}\,,
\end{equation}
\begin{equation}
 \label{generalphi}
\phi(t)=C(s)\left(-{V_{0}\over \gamma(s)}B(s)t^{3\over
{s+3}}+{D\over \Sigma_{0}}\right)^{-{{2s+3}\over
2s}}t^{-{(2s+3)^{2}\over {2s(s+3)}}}-\phi_{0}\,,
\end{equation}
where $D$ is the matter density constant, $\Sigma_{0}$ is a
constant of the motion resulting from the Noether symmetry,
$V_{0}$ is the constant that determines the scale of the
potential, $\phi_{0}$ is a constant that determines the initial
value of the scalar field and the other constants $A(s)$,
$B(s)$, $C(s)$, and $\gamma(s)$  are given in the appendix. As
it is apparent from Eq.(\ref{general}) and Eq.(\ref{generalphi})
for a generic value of $s$ both the scale factor $a(t)$ and the
scalar field $\phi(t)$ have a power law dependence on time. It
is also clear that there are two additional particular values of
$s$, namely $s=0$ and $s=-3$ which should be treated
independently. In this paper we concentrate on the case $s=-3$
which provides, as will be shown shortly, an interesting class
of models.
\subsubsection{The case of
quartic potentials: analysis of the solution } When $s=-3$  the
general solutions given by (\ref{general}) and
(\ref{generalphi}) lose their meaning. To find $a(t)$ and
$\phi(t)$ in this case it is necessary to use the general
procedure as described in (\cite{rug2}). From Eq.(\ref{e5}) and
Eq.(\ref{e6}) it follows that in this case  $
F=\frac{3}{32}\phi^{2}$, and $V(\phi)=V_0 \phi^{4}$ where we
have set $\phi_{0}=0$ and $V_0$ denotes a constant. This case is
particularly interesting since the resulting self--interaction
potential is used in finite temperature field theory. In fact,
it seems that the quartic form of the potential is required in
order to implement the symmetry restoration in several Grand
Unified Theories. Just for this reason we limit our analysis to
this special case, and we will show that it provides an
accelerated expansion of the universe. We reserve to a
forthcoming paper the study of the other cases. As in the
general case once we have the functions $F(\phi)$ and $V(\phi)$
that allow a Noether symmetry it is possible to explicitly find
it and effectively solve the Lagrange equations. Performing this
procedure we finally get
\begin{eqnarray}
  &&a(t) =\alpha_{0}e^{{-\alpha_1 t\over 3}}\left[{ \left(e^{\alpha_1\,t}-1 \right)\, +\alpha_2\,t +\alpha_3 }\right]^{\frac{2}{3}}\,,\\
   &&\phi(t) = \phi_0 \sqrt{e^{\alpha_1 t}\over  \left(e^{\alpha_1\,t} - 1\right) \, +\alpha_2\,t +\alpha_3
   }\label{eqaphi}\,,
\end{eqnarray}
\noindent where $\alpha_0$, $\alpha_1$, $\alpha_2$ $\alpha_3$
and $\phi_0$ are integration constants. They are related to the
initial matter density, $D$, and the scale of the potential
$V_0$ by  $D={{V_0} \over 16}\alpha_0^3\phi_0^{2}\alpha_1
\alpha_2$, which implies that they cannot be zero. The case
$V_0=0$ has to be treated separately. It turns out that the
constants $\alpha_3$, $\phi_0$ and $\alpha_0$ have an immediate
physical interpretation: $\alpha_0$, and $\alpha_3$ are
connected to the value of the scale factor at $t=0$, actually
$a(0)=(\alpha_3)^{2\over 3}\alpha_0$. Moreover $\alpha_3$ can be
selected in such a way that at a sufficiently early epoch the
universe is matter dominated. This requires that $\alpha_3$ be
sufficiently small, for example, that $\alpha_3\in
[0.001,0.01]$. The constants $\alpha_3$ and $\phi_0$ are
connected to the initial value of the scalar field
$\phi(0)=\displaystyle{\phi_0\over \sqrt{\alpha_3}}$ and
therefore they determine the initial value of the effective
gravitational constant
$G_{eff}(0)=-\displaystyle{{16\alpha_3}\over {3\phi_0^{2}}}$.
Let us note that an attractive gravity is recovered when $
\phi_0$ is a pure imaginary number. Without compromising the
general nature of the problem we can set, for example
$\phi_0=\imath$. This choice does not violate the positivity of
energy density of the scalar field or the weak energy condition.
To determine the integration constants $\alpha_1$, $\alpha_2$,
$\alpha_3$ and $\phi_0$  we follow the procedure used in
(\cite{pv}), and we set the present time $t_0 = 1$. That is to
say that we are using the age of the universe, $t_0$, as a unit
of time. Because of our choice of time unit the expansion rate
$H(t)$ is dimensionless, so that our Hubble constant is not the
same as the $H_0$ that appears in the standard FRW model,
measured in $km s^{-1}Mpc^{-1}$: we then set ${\widehat H}_0=
H(1)$. Using (29) we get
\[H(1)={\widehat H}_0=-{\alpha_1\over 3}+{2\over
3}{{{\alpha_1}e^{\alpha_1}+\alpha_2}\over {e^{\alpha_1}+\alpha_2+\alpha_3 - 1}}\,,\]
which we use to find $\alpha_2$ in the form
\[\alpha_2={{e^{\alpha_1}(\alpha_1-3{\widehat H}_0)+(\alpha_3 -
1)(3{\widehat H}_0+\alpha_1)}\over {3{\widehat
H}_0+\alpha_1-2}}\,.\] With this choice of time the scale factor,
the scalar field and the expansion rate assume the final form
 \begin{eqnarray}
  && a (t)=a_0e^{-\frac{\alpha_{1} t}{3}}\left(e^{\alpha_{1} t}
  +{{e^{\alpha_1}(\alpha_1-3{\widehat H}_0)+(\alpha_3 -
1)(3{\widehat H}_0+\alpha_1)}\over {3{\widehat
H}_0+\alpha_1-2}}t+\alpha_3 - 1\right)^{2/3}
\,,\label{scale}\nonumber
   \\ && \nonumber \\&&
   \phi(t)= \phi_0\sqrt{\frac{e^{\alpha_1 t}}{e^{\alpha_1 t}-{{e^{\alpha_1}(\alpha_1-3{\widehat H}_0)+(\alpha_3 -
1)(3{\widehat H}_0+\alpha_1)}\over {3{\widehat
H}_0+\alpha_1-2}}t+\alpha_3 -1}}\,,\label{fi}
    \end{eqnarray}
\begin{eqnarray}
  &&H(t)= \left\{\alpha_1 \left(\alpha_1+2
e^{\alpha_1}\right)+e^{\alpha_1 t} \alpha_1 \left(3
   {\widehat H}_0+\alpha_1-2\right)+ 3 {\widehat H}_0 \left(\alpha_1-2
   e^\alpha_1+2\right)\right.   \\
   &&\left. -\left(\alpha_1^2+3 {\widehat H}_0
   \left(\alpha_1+2\right)\right)
   \alpha_3 +\left[\left(-1-e^\alpha_1\right)
   \alpha_1^2+\alpha_3 \alpha_1^2+3
   \left(-1+e^\alpha_1+\alpha_3\right) {\widehat H}_0 \alpha_1\right] t\right\} \times \nonumber \\
   && \left\{\{3e^{\alpha_1 t} \left(3 {\widehat H}_0+\alpha_1-2\right)-\left(\alpha_3-1\right)
   \left(3 {\widehat H}_0 (t-1)+\alpha_1 (t-1)+2\right)+e^\alpha_1 (\alpha_1-3 {\widehat H}_0)
   t  \right\}^{-1}\nonumber\,.
\end{eqnarray}
 Let us remind that $t$ is now varying from $0$ to $1$ and $t=1$
corresponds to the present moment. The parameters ${\widehat
H}_0$ and $\alpha_1$ admit a simple physical interpretation.
Actually ${\widehat H}_0$ is the present value of the Hubble
constant measured in our unit of time, while $\alpha_1$ drives
the early time and the asymptotical behavior of $a(t)$ and
$\phi(t)$. For $t \ll {1\over \alpha_1 }$ we have
\begin{eqnarray}
  a(t) &\sim & \left[(\alpha_1+\alpha_2 - {\alpha_1\alpha_2\over 2})t+\alpha_3\right]^{2\over 3} \\
  \phi(t) &\sim & \left[(\alpha_1+\alpha_2 - \alpha_1\alpha_3)t+\alpha_3\right]^{- {1\over
  2}}.
\end{eqnarray}

Later at larger $t$, $a(t)$ reaches an intermediate stage, when
it evolves as  $a(t)\sim t^{2\over 3} e^{-{\alpha_1 t\over 3}}$
(dumped dust) , and has a de Sitter behavior $a(t)\sim
e^{{\alpha_1 t}\over 3}$ for $t\rightarrow \infty$.
\begin{figure}
\centering{\includegraphics[width=6 cm, height=5
cm]{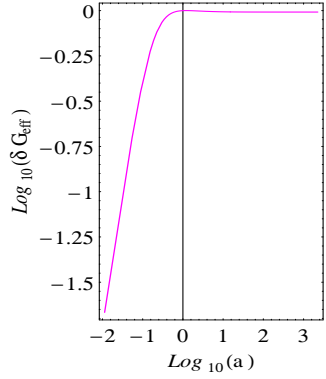}}
        \caption{\small  Plot of ${\log}_{10}\delta
G_{eff}=\displaystyle {\log}_{10}{G_{eff}\over G_{eff}(t_0)}$
versus ${\log}_{10} a$ . The vertical bar  marks ${\log}_{10}a_0$.
}
        \label{geff_def}
\end{figure}

\begin{figure}
\centering{\includegraphics[width=6 cm, height=5 cm]{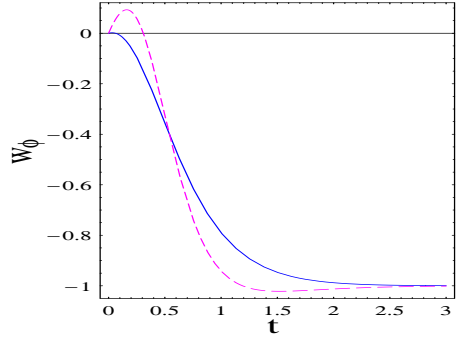}}
        \caption{\small Time dependence of  $w_\phi$,  for two
        different values of the parameters. Actually the solid
        curve corresponds to $\alpha_1= 2.5$, and ${\widehat H}_0 =0.95$, while the dashed
 one corresponds to $\alpha_1=3$,
        and ${\widehat H}_0 =1$. We see that even though they both {\it produce}
        accelerated expansion, only the second one gives rise to
        {\it super acceleration}. }
        \label{ww}
\end{figure}

\begin{figure}
\centering{\includegraphics[width=6 cm, height=5 cm]{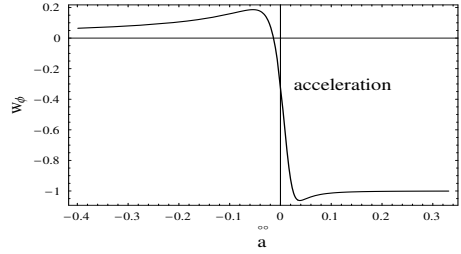}}
        \caption{\small Parametric plot of  $w_\phi$ as a function of the
        acceleration; we see the transition from the accelerated to the
        decelerated expansion.}
        \label{waccnew}
\end{figure}

\begin{figure}
\centering{\includegraphics[width=6 cm, height=5 cm]{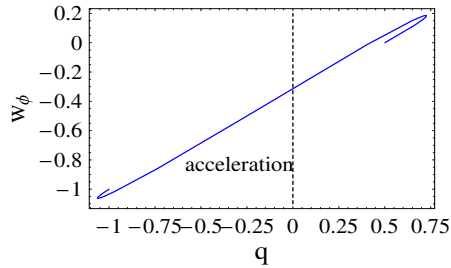}}
        \caption{\small Parametric plot of  $w_\phi$ with respect to the
        deceleration parameter $q=-{\ddot{a}\dot{a}\over a^2}$; again we see
        the transition from the accelerated to the decelerated expansion.
        Moreover, we see that if we consider the value $q\simeq -0.68$
        obtained from the SNIa Gold Sample (\cite{Riess04}) we obtain
        $w_\phi \leq -0.65$}
        \label{wacc2}
\end{figure}

\begin{figure}
\centering{
        \includegraphics[width=5 cm, height=4 cm]{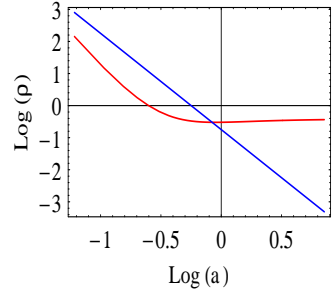}}
        \caption{\small Plot of ${\log}_{10}\rho_{\phi}$ versus ${\log}_{10} a$.
        The vertical bar marks ${\log}_{10}a_0$. The solid blue straight
        line indicates the log-log plot of $\rho_m$ versus a.}
        \label{nmc7ap_logrho}
\end{figure}

\begin{figure}
\centering{
        \includegraphics[width=5 cm, height=4 cm]{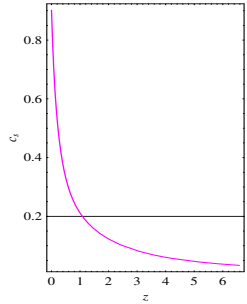}}
        \caption{\small The redshift behavior of the quintessence sound velocity $c_s$. }
        \label{sound}
\end{figure}

It is interesting to note that\footnote{Moreover, we will show
in the following that in order to fit the observational data
$\tilde{H}={\alpha_1\over 3}$ has to be of the same order of
magnitude as ${\widehat H}_0$, i.e. roughly $\alpha_1\approx 3
{\widehat H}_0$. This implies that this model even if formally
depending on two parameters is, as a matter of fact, very sturdy
and depends mainly on the Hubble constant. } $w_{\phi}$ is
representing an equation of state, in the usual sense, of the
effective cosmological constant $\Lambda_{eff.}$. In
Figs.(\ref{ww},\ref{waccnew},\ref{wacc2}) we show the time
dependence of $w_{\phi}$: we see that this equation of state can
{\it admit a superquintessence behavior} ($w<-1$), just as an
effect of the transition toward $w\rightarrow -1$.
We note that in the remote past as in the far future $w_{\phi}$ is
constant: it mimics an {\it almost} dust equation of state
($w_{\phi}\approx 0$) in the far past and asymptotically behaves as
a bare cosmological constant ($w_{\phi}\rightarrow -1$) as $t
\rightarrow \infty$. Let us also mention that since both
$\rho_{\phi}$ and $p_\phi$ depend on $F(\phi)$ through its time
derivative and asymptotically $\phi(t)\sim {\rm constant}$ we
asymptotically {\it recover} the minimally coupled case, with
\begin{eqnarray}\label{minasymp}
  \rho_{\phi_{\infty}} &=&  \frac {1}{2} {{\dot \phi}_{\infty}}^2+ V_{\infty}(\phi) =
  V_0\phi_{0}^{4}+{1\over 8}\phi_{0}^{2}e^{-2\alpha_1 t}\,,\\
 p_{\phi_{\infty} }&=&  \frac {1}{2} {{\dot \phi}_{\infty}}^2-
 V_{\infty}(\phi)= -V_0\phi_{0}^{4}+{1\over 8}\phi_{0}^{2}e^{-2\alpha_1 t}\,.
\end{eqnarray}
However, before reaching this asymptotic regime the total
energy density $\rho_\phi$ is  dominated by the coupling term
$6H\dot{F}$.
In conclusion of this section we present the traditional plot
$\log{\rho_{\phi}}$ - $\log a$ compared with the matter density
(see Fig. (\ref{nmc7ap_logrho})). Interestingly we see that
$\rho_{\phi}$ tracks the matter during the matter dominated era
( actually $a(t)\propto t^{2\over 3}$, and $w_{\phi}\sim 0$),
and becomes dominant at late time. In Fig.(\ref{sound}) we plot
the redshift behavior of the quintessence sound velocity; we see
that $c_s< 1$.
\subsubsection{A special case: asymptotic freedom at   $t\simeq 0$  }
\begin{figure}
\centering{
        \includegraphics[width=6 cm, height=4 cm]{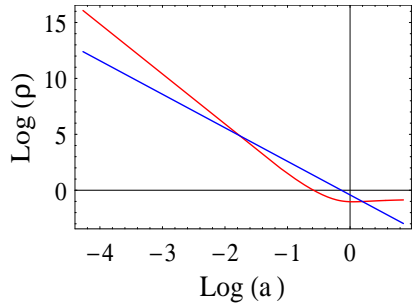}}
        \caption{\small Plot of ${\log}_{10}\rho_\phi$ versus ${\log}_{10} a$
        when there is asymptotic freedom at   $t\simeq 0$ .
        The vertical line marks ${\log}_{10}a_0$. The solid blue straight
        line indicates the log-log plot of $\rho_m$ versus a.}
        \label{rhomatterzero}
\end{figure}

\begin{figure}
\centering{
        \includegraphics[width=7 cm, height=7 cm]{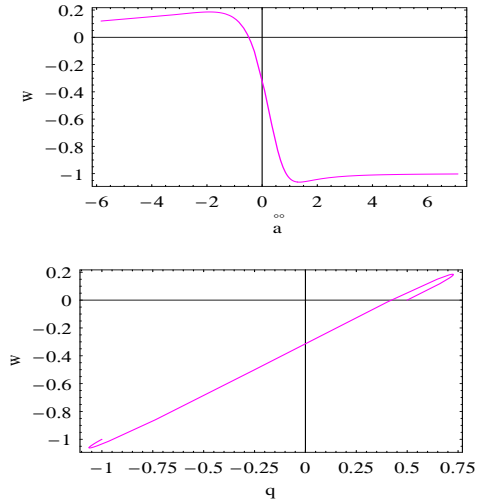}}
        \caption{\small The same as in Figs (\ref{waccnew},\ref{wacc2})when there is
        asymptotic freedom at   $t\simeq 0$ .}
        \label{wqaccay}
\end{figure}

In this section we consider a special case when
$\lim_{t\rightarrow 0} G_{eff}=0 $, that is when
$\lim_{t\rightarrow 0} \phi = \infty$ we have a sort of {\it
asymptotic freedom } at $t=0$. First of all we note that such a
case can be reached by setting $\alpha_3=0$, which also implies
that $a(t=0)=0$, and the expressions for $a(t)$ and $\phi(t)$
become simpler. In comparison with  the case $\alpha_3\neq 0$
the main difference concerns the behavior of the density
$\rho_{\phi}$ with respect to the matter density, as shown in
Fig. (\ref{rhomatterzero}). Actually we note that the coupling
$F(\phi)$ diverges as $t\rightarrow 0$, with its derivatives,
more rapidly than $\rho_m$. However $\rho_m$ always dominates
over the scalar field contribution to the density, that is over
$\rho_\phi$.
It should be remembered that the early epoch $t\approx 0$ does not
belong  to the physical time domain of our model, since we are
neglecting the contribution of radiation, and therefore the time
behavior of $G_{eff}$ during the intermediate period is unknown.
This is also the reason why we are not using any constraint on
$G_{eff}$ from the nucleosynthesis. With respect to the other
characteristic features this case is similar to that discussed
above, as shown in the Fig. (\ref{wqaccay})
\section{Observational data and predictions of our models}
\begin{figure}
\centering{
\includegraphics[width=6 cm, height=6. cm]{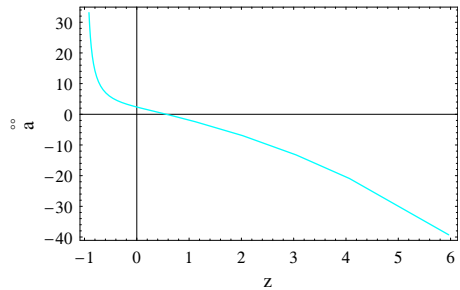}}
 \caption{\small Behavior  of the second derivative of the scale
   factor. Please note that the transition from a decelerating to an
accelerating expansion occurs close to $z= 0.5 $,
 as predicted by recent observations of SNIa $z_t= 0.46\pm 0.13$ (\cite{Riess04}).}
\label{transition}
\end{figure}

Above we discussed some general properties of our scalar field model
of quintessence, stressing how it provides a {\it natural mechanism}
for the observed accelerated expansion of the universe. To test
viability of our model we compare its predictions with the available
observational data. We concentrate mainly on two different kinds of
observational data: some of them  are all based on distance
measurements, as the publicly available data on type Ia supernovae,
the measurements of cosmological distances with the
Sunyaev-Zel'dovich effect and radio-galaxies data, other are instead
connected with the large scale structure, such as the parameters of
large scale structure determined by the 2-degree Field Galaxy
Redshift Survey (2dFGRS), and the gas fraction in clusters. Before
we begin our analysis we note in Fig. \ref{transition} that for our
model the transition redshift from a decelerating to an accelerating
phase in the evolution of the universe falls very close to $z=0.5$,
in agreement with recent results coming from the SNIa observations
(\cite{Riess04}).\begin{figure}[tbp]
\begin{minipage}{30 cm}
\begin{minipage}{7 cm}
 \includegraphics[width=6 cm, height=7 cm]{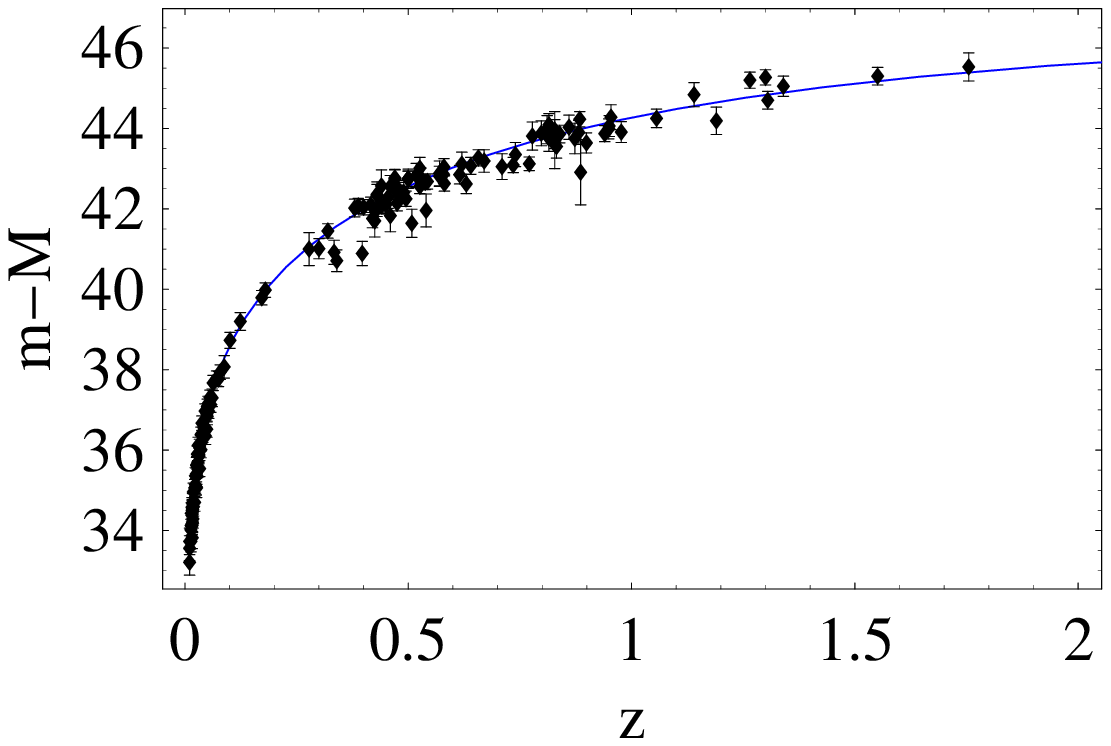}
        \caption{\small Observational data of the Gold Sample of
    SNIa (Riess et al. 2004) and the SNLS dataset (\cite{SNLS}) fitted to our model. The
    solid curve is the best fit curve with
    ${\widehat H}_0=1.0^{+0.03}_{-0.04}$, $\alpha_1=2.9^{+0.2}_{-0.3}$,
    which corresponds to $\Omega_{{\rm \Lambda_{eff}}}=0.84_{-0.07}^{+0.06}$.
    We also get $h=0.68^{+.05}_{-.03}$.  }
        \label{sup}
\end{minipage}
\hspace*{50 pt}
\begin{minipage}{8 cm}
 \includegraphics[width=6 cm, height=7 cm]{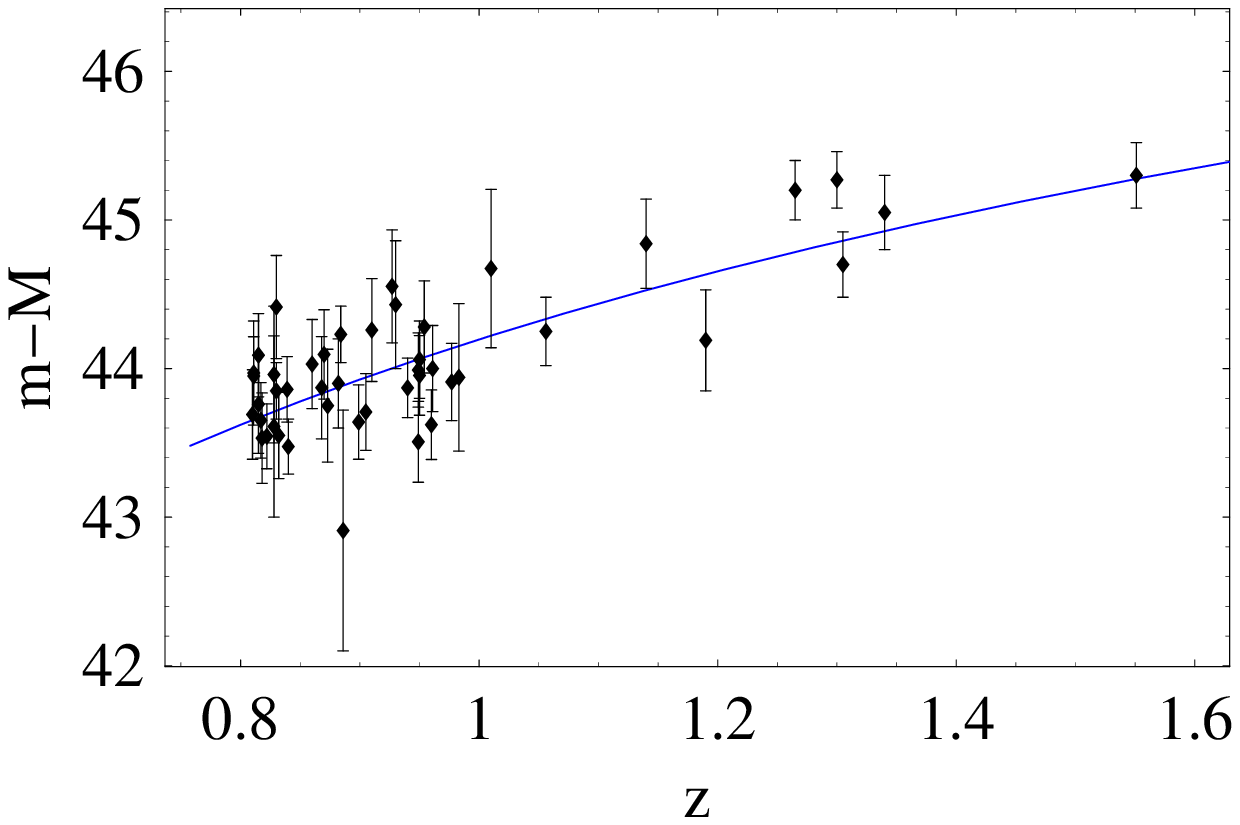}
        \caption{\small The same as in the Eq. (\ref{sup}) , but zooming on the high
redshift SNIa.   }
        \label{hubblesn+gr2}
\end{minipage}
\end{minipage}\\
\end{figure}
\subsection{Constraints from recent SNIa observations}
In recent years the confidence in type Ia supernovae as standard
candles has been steadily growing. Actually it was just the SNIa
observations that gave the first strong indication of an
accelerating expansion of the universe, which can be explained by
assuming the existence of some kind of dark energy or nonzero
cosmological constant (\cite{Schmidt}). Since 1995 two teams of
astronomers have been discovering type Ia supernovae at high
redshifts. First results of both teams were published by Schmidt
\& al. (1998) and Perlmutter \& al. (1999). Recently the High-Z SN
Search Team reported discovery of 8 new supernovae in the redshift
interval $0.3\leq z \leq 1.2$ and they compiled data on 230
previously discovered type Ia supernovae (\cite{Tonry}). Later
Barris \& al. (2004) announced the discovery of twenty-three
high-redshift supernovae spanning the  range of $z=0.34 - 1.03$,
including 15 SNIa at $z\geq 0.7$~.\\ More recently  Riess \& al.
~(2004) announced the discovery of 16 type Ia supernovae with the
Hubble Space Telescope. This new sample includes 6 of the 7 most
distant ($z> 1.25$) type Ia supernovae. They determined the
luminosity distance to these supernovae and to 170 previously
reported ones using the same set of algorithms, obtaining in this
way a uniform "Gold Sample" of type Ia supernovae containing 157
objects. Finally just recently the Supernova Legacy Survey (SNLS)
team presented the data collected during the first year of the
SNLS program. It consists of 71 high redshift supernovae in the
redshift range $ z \in \left[0.2, \,\,1\right]$. This new SNLS
sample is characterized by precise distance measurements of all
the $71$ supernovae, so it can be used to build the Hubble diagram
extending to $z=1$. The purpose of this section is to test our
scalar field quintessence model by using the best SNIa data sets
presently available. As a starting point we consider the gold
sample compiled  by (\cite{Riess04}) to which we add the SNLS
dataset. To constrain our model we compare through a $\chi^2$
analysis the redshift dependence of the observational estimates of
the distance modulus, $\mu=m-M$, to the corresponding theoretical
values. The distance modulus is generally defined by
\begin{equation}
m-M=5\log{D_{L}(z)}+5\log({c\over H_{0}})+25, \label{eq:mMr}
\end{equation}
where  $H_0$ is the standard Hubble constant, measured in $km
s^{-1}Mpc^{-1}$, $m$ is the appropriately corrected apparent
magnitude including reddening, K correction etc., $M$ is the
corresponding absolute magnitude, and $D_{L}$ is the luminosity
distance in Mpc. However, in scalar tensor theories of gravity
it is important also to include in the Eq. (\ref{eq:mMr})
corrections, which describe the effect of the time variation of
the effective gravitational constant $G_{eff}$ on the luminosity
of high redshift supernovae. Actually, if the local value of
$G_{eff}$ at the space time position of the most distant
supernovae differs from $G_N$, this could in principle induce a
change in the Chandrasekhar mass $M_{ch}\propto G^{-{3\over
2}}$.
 Some analytical models of the
supernovae light curves predict that the peak luminosity is
proportional to the mass of nickel produced during the
explosion, which is a fraction of the Chandrasekhar mass. The
actual fraction varies in different scenarios, but always the
physical mechanism of type Ia supernovae explosion relates the
energy yield to the Chadrasehkar mass. Assuming that the same
mechanism for the ignition and the propagation of the burning
front is valid for SNIa at high and low redshifts, it turns out
that the predicted apparent magnitude will be fainter by a
quantity (\cite{Gat01})
\begin{equation}
\Delta M_{G}={15\over 4}\log\left(G_{eff}\over
G_{eff_0}\right)\label{corgef}.
\end{equation}
Taking this into account the distance modulus becomes
\begin{equation}
m-M=5\log{D_{L}(z)}+5\log({c\over H_{0}})+25 + \Delta M_{G}.
\label{eq:modg}
\end{equation}
The presence of this correction actually allows one to test the
scalar tensor theories of gravity (\cite{Gat01,uz}) using the
SNIa data. For a general flat and homogeneous cosmological model
the luminosity distance can be expressed as an integral of the
Hubble function as follows:
\begin{eqnarray}\label{luminosity}
D_L (z) &=& {c \over H_0}(1+z)\int^{z}_{0}{1\over H(\zeta)}d\zeta,
\end{eqnarray}
where $H(z)$ is the Hubble function expressed in terms of
$z=a_0/a(t) - 1$. Using Eqs. (\ref{corgef}) and
(\ref{luminosity}), which in our case can be integrated only
numerically, we construct the distance modulus and perform the
$\chi^2$ analysis on the complete data set. We obtain
$\chi_{red}^2=1.7$ for 230 data points, and the best fit value
is  ${\widehat H}_0=1.0^{+0.03}_{-0.04}$,
$\alpha_1=2.9^{+0.2}_{-0.3}$,
 which corresponds to $\Omega_{{\rm \Lambda_{eff}}}=0.73_{-0.07}^{+0.06}$. We also get $h=0.68^{+.05}_{-.03}$.
Moreover we checked that any value of $\alpha_3\in [0.001,
0.01]$ does not affect the determination of distances, so that
in the following we set $\alpha_3=0.001$, being confident that
such a choice does not alter the main results. In Fig
(\ref{sup}) we compare the best fit curve with the observational
data sets.
\subsubsection{Dimensionless
coordinate distance test}
\begin{figure}
\centering{
        \includegraphics[width=6 cm, height=6 cm]{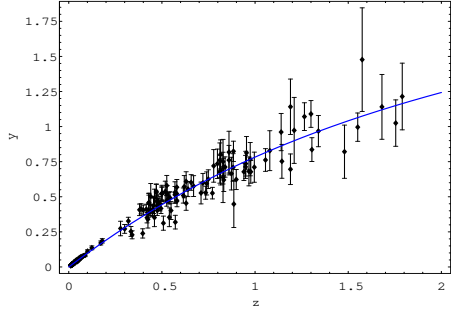}}
        \caption{\small Observational Daly \& Djorgovski database (\cite{DD04})
        fitted to our model. The solid curve is the best fit curve with
        $\chi_{red}^2=1.19$ for 186 data points,  and the best
        fit value is ${\widehat H}_0=1.00^{+0.05}_{-0.03}$,
        $\alpha_1=2.5^{+0.3}_{-0.2}$. }
        \label{cord}
\end{figure}

\begin{figure}
\centering{
        \includegraphics[width=7cm, height=6 cm]{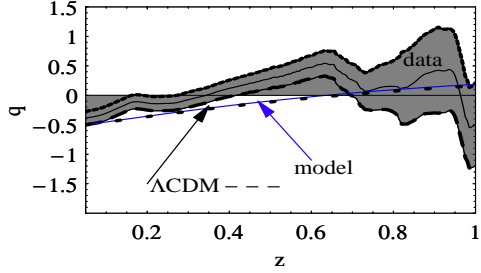}}
        \caption{\small A first look at the allowed region for $q(z)$,
        obtained by Daly \& Djorgovski
        from the full data set (shadow area). Approximated
        polynomial
        corresponding to a $z$-window $\Delta z =
        0.6$ is shown with the black thin solid line, with the black thick dashed lines
        are shown the
        approximated polynomial fitted to the smoothed data  at
        $\pm 1\sigma$ range, and corresponding to a $z$-window $\Delta z = 0.4$.
        The blue solid line
        shows the deceleration function, q(z), for our model
        corresponding to the the best fit values  ${\widehat H}_0=1.00^{+0.05}_{-0.03}$,
        $\alpha_1=2.5^{+0.3}_{-0.2}$.
        It is compared to the corresponding function for the {\it standard} $\Lambda$CDM
        model with $\Omega_M=0.3$, $\Omega_{\Lambda}=0.7$. }
        \label{dalyq}
\end{figure}

After having explored the Hubble diagram of SNIa, that is the
plot of the distance modulus as a function of the redshift $z$,
we want here to follow a very similar, but more general
approach, considering as cosmological observable the
dimensionless coordinate distance defined as\,:
\begin{equation}
y(z) = \int_{0}^{z}{\frac{1}{H(\zeta)} d\zeta} \ . \label{eq:
defy}
\end{equation}
 It is worth noting that $y(z)$ does not depend explicitly on
$h$ so that any choice for $h$ does not alter the main result.
Daly \& Djorgovski (\cite{DD04}) have determined $y(z)$ for the SNIa in the Gold
Sample of Riess et al. (\cite{Riess04}) which represents the most
homogeneous SNIa sample available today. Since SNIa allows to estimate $D_L$
rather than $y$, a value of $h$ has to be set. Fitting the Hubble
law to a large set of low redshift ($z < 0.1$) SNIa, Daly \& Djorgovski
(\cite{DD04}) have found that\,:
\begin{displaymath}
h = 0. 66 \pm 0.08 \ {\rm km \ s^{-1} \ Mpc^{-1}} \,,
\end{displaymath}
which is consistent with our fitted value
$h=0.66^{+.05}_{-.03}$. To enlarge the sample, Daly \&
Djorgovski added 20 further points on the $y(z)$ diagram using a
technique of distance determination based on the angular
dimension of radiogalaxies (\cite{DD04}). This extended sample
that spans the redshift range $(0.1, 1.8)$   has been obtained
by homogenizing different kinds of measurements, affected by
different systematics, so that the full sample may be used
without introducing spurious features in the $y(z)$ diagram.
However before using the dimensionless coordinate distance test
we do not use the Daly \& Djorgovski database directly, but, for
the supernovae Gold Sample we first converted the distance
modulus into $y(z_i)$ using the Eq. (\ref{eq:modg}) and the
relation
\begin{displaymath}
D_L = \frac{c}{H_0} (1 + z) y(z).
\end{displaymath}
Here again $H_0 $ is the standard FRW Hubble constant. To
determine the best fit parameters, we define the following merit
function\,:
\begin{equation}
\chi^2(\alpha_1, {\widehat H}_0) = \frac{1}{N - 3} \sum_{i =
1}^{N}{\left [ \frac{y(z_i;  \alpha_1, {\widehat H}_0) -
y_i}{\sigma_i} \right ]^2}\,. \label{eq: defchi}
\end{equation}
We obtain $\chi_{red}^2=1.19$ for 186 data points,  and the best fit
value is ${\widehat H}_0=0.98^{+0.05}_{-0.03}$,
$\alpha_1=2.5^{+0.3}_{-0.2}$. In Fig (\ref{cord})  we compare the
best fit curve with the observational data set.
 Daly \& Djorgovski (\cite{DD04}) developed a numerical method for a direct determination of the expansion and acceleration rates, $H(z)$ and $q(z)$,
from the data, just using the dimensionless coordinate distance $y(z)$, without making any assumptions about the nature or evolution of the dark
energy. They actually use the equation
\begin{equation}
-q(z) \equiv \ddot{a} a/\dot{a}^2 = 1~ +~ (1+z) ~(dy/dz)^{-1}
(d^2y/dz^2)\,,\label{dalyeq}
\end{equation}
valid for $k=0$. Equation (\ref{dalyeq}) depends only upon the
Friedman-Robertson-Walker line element and the relation $(1+z) =
a_0/a(t)$. Thus, this expression for $q(z)$ is valid for any
homogeneous and isotropic universe in which $(1+z) = a_0/a(t)$, and
is therefore quite general and can be compared with any model to
account for the acceleration of the universe. This new approach has
the advantage of being model independent, but it introduces larger
errors in the estimation of $q(z)$, because the numerical derivation
is very sensitive to the size and quality of the data. An additional
problem is posed by the sparse and not complete coverage of the
$z$-range of interest. Measurement errors are propagated in the
standard way leading to estimated uncertainties of the fitted
values. In Fig. (\ref{dalyq}) we compare the $q(z)$ obtained by Daly
\& Djorgovski from their full data set with our {\it best fit}
model.
\subsection{The Sunyaev-Zeldovich/X-ray method}
\begin{figure}
\centering{
        \includegraphics[width=6 cm, height=6cm]{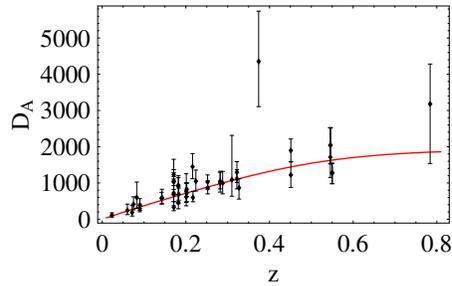}}
        \caption{\small Observational SZE data fitted to our model with
         the best fit values ${\widehat H}_0=0.97^{+0.04}_{-0.03}$,
$\alpha_1=3.2^{+0.1}_{-0.1}$, and $h=0.75\pm 0.05$} .
        \label{szall}
\end{figure}

In this section we discuss how the parameters of our model can
be also constrained by the angular diameter distance $D_A$ as
measured using the Sunyaev-Zeldovich effect (SZE) and the
thermal bremsstrahlung (X-ray brightness data) for galaxy
clusters. Actually in a  homogenous and isotropic cosmological
model the angular diameter distance can be easily related to the
coordinate distance leading to
\begin{eqnarray}
  D_A &=& {c\over H_0} {1\over (1+z)}\int_{0}^{z}{\frac{1}{H(\zeta)}
  d\zeta}.
\end{eqnarray}
 Distance measurements using SZE
and X-ray emission from the intracluster medium are based on the fact
that these processes depend on different combinations of some
parameters of the clusters (see \cite{birk} and references
therein). The SZE is a result of the inverse Compton scattering of the
CMB photons on hot electrons of the intracluster gas. The number of
photons is preserved, but photons gain energy and thus a decrement of
the temperature is generated in the Rayleigh-Jeans part of the
black-body spectrum while an increment appears in the Wien region.  We
limit our analysis to the so called {\it thermal} or {\it static} SZE,
which is present in all the clusters, neglecting the ${\it kinematic}$
effect, which is present only in clusters with a nonzero peculiar
velocity with respect to the Hubble flow along the line of
sight. Typically the thermal SZE is an order of magnitude larger than
the kinematic one. The shift of temperature is:
\begin{equation}
\frac{\Delta T}{T_0} = y\left[ x \, \mbox{coth}\,
\left(\frac{x}{2} \right) -4 \right], \label{eq:sze5}
\end{equation}
where ${\displaystyle x=\frac{h \nu}{k_B T}}$ is a dimensionless
variable, $T$ is the {\it shifted} radiation temperature, $T_0$
is the unperturbed CMB temperature and $y$ is the so called
Compton parameter, defined as the optical depth $\tau = \sigma_T
\int n_e dl$ times the energy gain per scattering:
\begin{equation}\label{compt}
  y=\int  \frac{k_B T_e}{m_e c^2} n_e \sigma_T dl.
\end{equation}
In  Eq.~(\ref{compt}), $T_e$ is the temperature of the electrons
in the intracluster gas, $m_e$ is the electron mass, $n_e$ is the
number density of the electrons, and $\sigma_T$ is the cross
section of Thompson electron scattering.  We have used the
condition $T_e \gg T_0$ ($T_e$ is of the order of $10^7\, K$ and
$T_0$,  is the CMB temperature $\simeq 2.7K$) and we assumed that
the CMB temperature varies linearly with redshift what implies
that after recombination the CMB radiation cools adiabatically
with no injection of energy in the form of photons. In the low
frequency regime of the Rayleigh-Jeans approximation we obtain
\begin{equation}
 \frac{\Delta T_{RJ}}{T_0}\simeq -2y\,.
 \label{eq:sze5bis}
\end{equation}
The next step  to quantify the SZE decrement is to specify the
model for the intracluster electron density and temperature
distribution. The most commonly used model is the so called
isothermal $\beta$ model of \cite{cavaliere}. In this model
\begin{eqnarray}
& &  n_e (r) = n_{e_0} \left( 1 + \left(
\frac{r}{r_e} \right)^2 \right)^{-\frac{3 \beta}{2}}\,, \\
& & T_e (r) = T_{e_0}~, \label{eq:sze6}
\end{eqnarray}
where $n_{e_0}$ and $T_{e_0}$ are respectively the central
electron number density and temperature of the  intracluster
electron gas, $r_e$ and $\beta$ are fitting parameters connected
with the model (\cite{sarazin}). The relative temperature shift is
given by
\begin{equation}
\frac{\Delta T}{T_0} = -\frac{2 k_B \sigma_T T_{e_0} \, n_{e_0}}{m_e
c^2} {\cdot} \Sigma \,,\label{eq:sze7}
\end{equation}
where
\begin{equation}
\Sigma = \int^\infty_0 \left( 1 + \left( \frac{r}{r_c} \right)^2\right)^{-\frac{3 \beta}{2}} dl\, , \label{eq:sze8}
\end{equation}
which depends only on the geometry and the extension of the cluster along the line of sight. In Eq.(\ref{eq:sze8}), $l$ is the coordinate along
the line of sight, $r^2=l^2+R^2$, and $R^2=x^2+y^2$. A simple geometrical argument converts the integral in Eq.(\ref{eq:sze8}) into an angular
form. Introducing the angular diameter distance, $d_A$, to the cluster we can rewrite (\ref{eq:sze7}) as
\begin{equation}\label{eq:sze10bis}
\frac{\Delta T (\theta =0)}{T_0}=- 2\frac{\sigma_T k_B T_{e}n_{e_0}}{m_e} \sqrt{\pi} \frac{\Gamma \left( \frac{3 \beta}{4}\right)}{\Gamma \left(
\frac{3\beta}{2}\right)}\frac{c}{H_0} d_A,
\end{equation}
where $T_e$ is the gas temperature. The factor $\displaystyle \frac{c}{H_0} d_A$ in Eq.~(\ref{eq:sze10bis}) carries the dependence of the thermal
SZE on the cosmological models (for a discussion of the dependence of $d_A$ on the {\it standard} $\Lambda$CDM model see (\cite{app})). From
Eq.~(\ref{eq:sze10bis}), we also note that the central electron number density is proportional to the inverse of the angular diameter distance,
actually
\begin{equation}
\label{frac4} n_{\rm e0}^{\rm SZ} \propto \frac{\Delta T_{\rm
SZ}}{T_{0}}\frac{1}{d_{\rm A}}.
\end{equation}
>From an independent point of view the central number density of
electrons can be also measured by fitting the X-ray surface
brightness profile, $S_{\rm X} \propto \int n_{\rm e}^2 \Lambda
(T_{\rm e})dl$, where the integration is along the line of sight
and $\Lambda (T_{\rm e})$ is the X-ray emissivity at the electron
temperature $T_{\rm e}$. It turns out that
\begin{equation}
\label{frac3} n_{\rm e0}^X \propto \sqrt{\frac{S_{\rm X}}{d_{\rm A}}}.
\end{equation}
By eliminating $n_{\rm e0}$ from Eqs.~(\ref{frac3}), and
~(\ref{frac4}), one can solve for the angular diameter distance,
yielding
\begin{equation}
\label{frac7} d_{\rm A} \propto \frac{(\Delta T_{\rm SZ})^2}{S_{\rm X}}.
\end{equation}
Recently distances to 18 clusters with redshift ranging from
$z\sim 0.14$ to $z\sim 0.78$ have been determined from a
likelihood joint analysis of SZE and X-ray observations (see Table
7 in \cite{reese}). We perform our analysis using angular diameter
distance measurements for a sample of 44 clusters, containing the
18 above  mentioned clusters and other 24 known previously (see
\cite{birk}). We perform a statistical analysis on the SZE data
defining the following merit function\,:
\begin{eqnarray}
\label{eq: defchitot}
 &&\chi^2(\alpha_1, {\widehat H}_0) = \frac{1}{M - 3} \sum_{i =
1}^{M}\left[ \frac{(D_A(z_i; \alpha_1, H0)-D_i) }{\sigma_i}
\right]^2.
\end{eqnarray}
 We obtain $\chi_{red}^2=1.14$ for  44 data points, and the
best fit values are ${\widehat H}_0=0.97^{+0.04}_{-0.03}$,
$\alpha_1=3.2^{+0.1}_{-0.1}$. We also get $h=0.75\pm 0.05$. In
Fig. (\ref{szall}) we compare the best fit curve with the
observational SZE data.

\subsection{Gamma-Ray Burst Hubble Diagram }
\begin{figure}
\centering{
        \includegraphics[width=6 cm, height=6 cm]{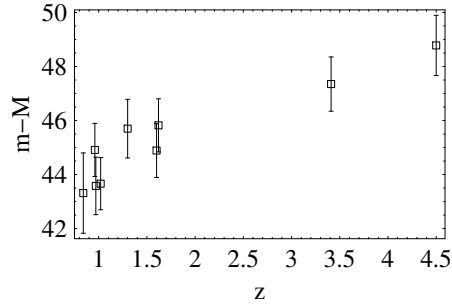}}
        \caption{\small Hubble diagram for the  BATSE  gamma ray bursts
        (\cite{schaefer}) up to $z=4.5$.  }
        \label{grb1}
\end{figure}

\begin{figure}[tbp]
\begin{minipage}{30 cm}
\begin{minipage}{7 cm}
 \includegraphics[width=6 cm, height=7 cm]{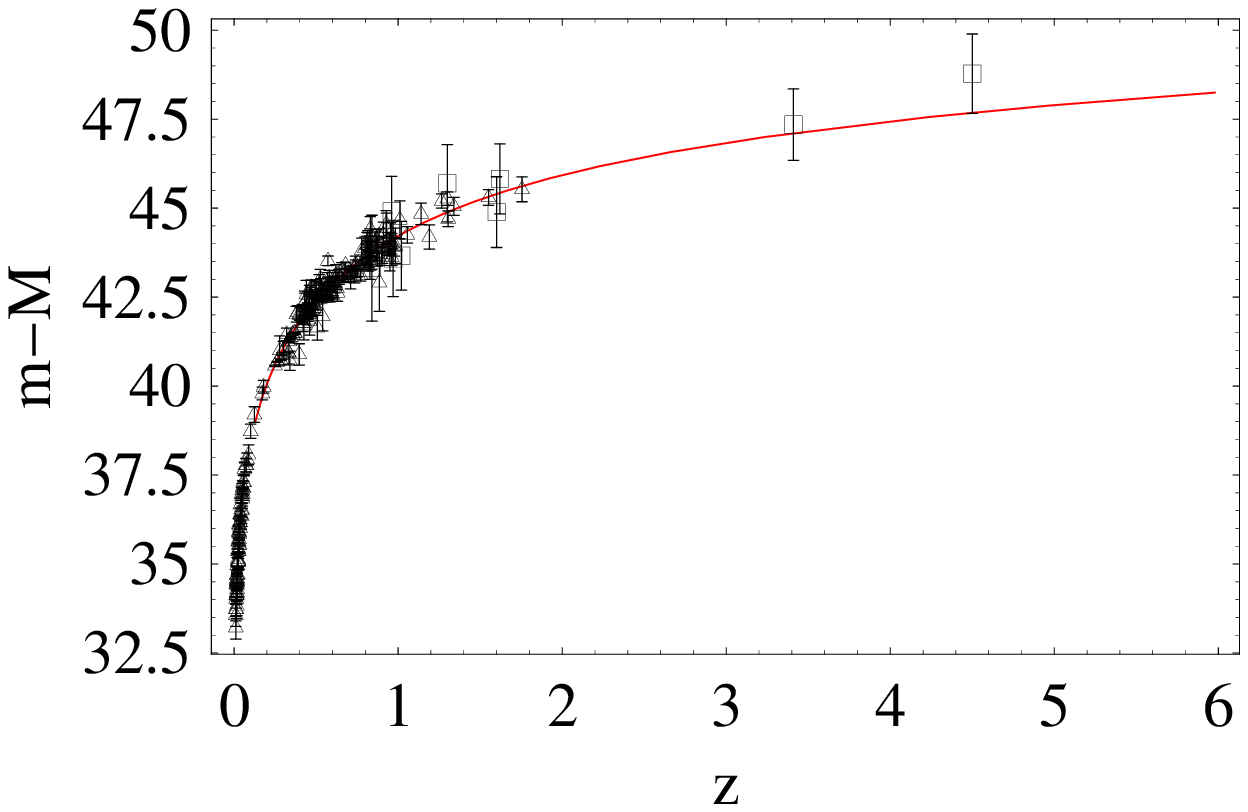}
        \caption{\small Observational Hubble diagram for the SNIa Gold Sample
        (Riess et al. 2004), the SNLS data(\cite{SNLS}) (filled boxes), and the  BATSE  GRBs
         data (\cite{schaefer}) (empty boxes) fitted to our model. The
         solid curve is the best fit curve with ${\widehat H}_0=1^{+0.05}_{-0.04}$,
         $\alpha_1=2.8^{+0.1}_{-0.2}$, and $h=0.68\pm 0.05$. }
        \label{hubblesn+grb}
\end{minipage}
\hspace*{50 pt}
\begin{minipage}{8 cm}
 \includegraphics[width=6 cm, height=7 cm]{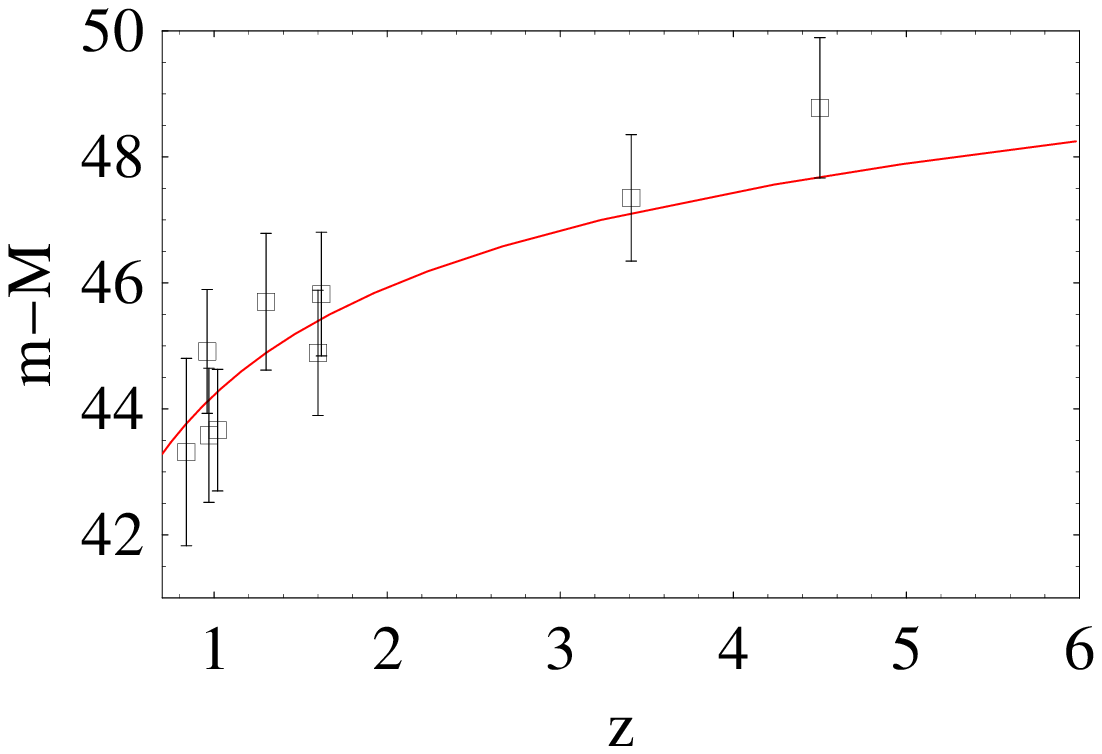}
        \caption{\small The same as in the Eq. (\ref{hubblesn+grb}) ,
      but zooming on the high redshift GRBs.   }
        \label{hubblesn+gr2b}
\end{minipage}
\end{minipage}\\
\end{figure}

Gamma-ray bursts (GRBs) are bright  explosions visible across most
of the Universe, certainly out to redshifts of $z=4.5$ and likely
out to $z \sim 10$. Recent studies have pointed out that GRBs may
be used as standard cosmological candles. The prompt energy
released during burst spans nearly three orders of magnitude, and
the distribution of the opening angles of the emission, as deduced
from the timing of the achromatic steepening of the afterglow
emission, spans a similar wide range of values. However, when the
apparently isotropic energy release and the conic opening of the
emission are combined to infer the intrinsic, true energy release,
the resulting distribution does not widen, as is expected for
uncorrelated data, but shrinks to a very well determined value
(\cite{fk03}), with a remarkably small (one--sided) scattering,
corresponding to about a factor of $2$ in total energy. Similar
studies in the X--ray band have reproduced the same results. It is
thus very tempting to study to what extent this property of GRBs
makes them suitable cosmological standard candles. Schaefer
(\cite{schaefer}) proposed using two well known correlations of
the GRBs luminosity (with variability, and with time delay) to the
same end, while other exploited the recently reported relationship
between the beaming--corrected $\gamma$-ray energy and the locally
observed peak energy of GRBs (see for instance \cite{day}). As for
the possible variation of ambient density from burst to burst,
which may widen the distribution of bursts energies, Frail \&
Kulkarni (\cite{fk03}) remarked that this spread is already
contained in their data sample, and yet the distribution of energy
released is still very narrow. There are at least two reasons  why
GRBs are better than type Ia supernovae  as cosmological candles.
On the one hand, GRBs are easy to find and locate: even  1980s
technology allowed BATSE to locate $\sim$1 GRB per day, despite an
incompleteness of about $1/3$, making the build--up of a
300--object database a one--year enterprise. The {\it Swift}
satellite launched on 20 November 2004, is expected to detect GRBs
at about the same rate as BATSE, but with a nearly perfect
capacity for identifying their redshifts simultaneously with the
afterglow observations
\footnote{http://swift.gsfc.nasa.gov/docs/swift/proposals/appendix\_f.html}.
Second, GRBs have been detected out to very high redshifts: even
the current sample of about 40 objects contains several events
with $z> 3$, with one (GRB 000131) at $z = 4.5$. This should be
contrasted with the difficulty of locating SN at  $z
> 1$, and the absolute lack of any SN with $z
> 2$.
On the other hand, the distribution of luminosities of SNIa is
narrower than the distribution of energy released by GRBs,
corresponding to a magnitude dispersion $\sigma_M = 0.18$ rather
than $\sigma_M = 0.75$. Thus GRBs may provide a complementary
standard candle, out to distances which cannot be probed by SNIa,
their major limitation being the larger intrinsic scatter of the
energy released, as compared to the small scatter in peak
luminosities of SNIa. There currently exists enough information to
calibrate luminosity distances and independent redshifts for nine
bursts (\cite{schaefer}). These bursts were all detected by BATSE
with redshifts measured from optical spectra of either the
afterglow or the host galaxy.  The highly unusual GRB980425
(associated with supernova SN1998bw) is not included because it is
likely to be qualitatively different from the classical GRBs.
Bursts with red shifts that were not recorded by BATSE cannot yet
have their observed parameters converted to energies and fluxes
that are comparable with BATSE data. We perform our analysis using
the data shown in  Fig. (\ref{grb1})  with the distance modulus
$\mu$, given by Eq. (\ref{eq:modg}).
To this aim, the only difference with respect to the SNIa is
that we {\it slightly} modify the {\it correction term} of Eq.
(\ref{corgef}), into
\begin{equation}\label{corgef2}
\Delta m_{G_{eff}}=2.5\gamma \frac{\Delta G_{eff}(t)}{\left( \ln
10\right) G_{eff}}.
\end{equation}
We expect that $\gamma $ is of order unity, so that the
$G$-correction would be roughly half a magnitude. We obtain
$\chi_{red}^2=1.09$, and the best fit value is ${\widehat
H}_0=1^{+0.05}_{-0.04}$, $\alpha_1=2.8^{+0.1}_{-0.2}$, and
$h=0.66\pm 0.05$, which are compatible with the SNIa results. We
also confirm that $\gamma=1.5$ as in Eq. (\ref{corgef}). In Fig.
(\ref{hubblesn+grb}) we compare the best fit curve with both the
GRBs and the SNIa Gold Sample.

\subsection{The gas fraction in clusters}
\begin{figure}
\centering{
        \includegraphics[width=8 cm, height=10 cm]{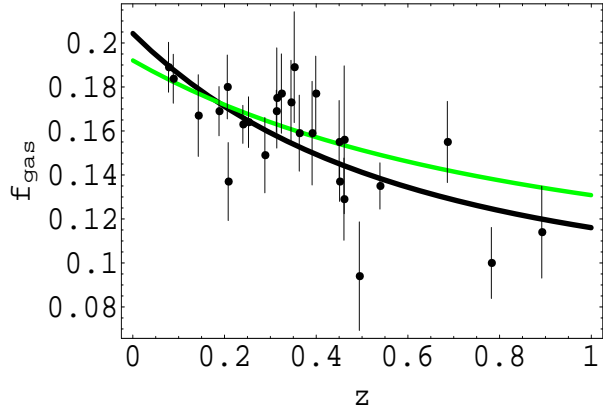}}
        \caption{\small In the diagram we plot the best fit curve to the $f_{gas}$
        data for our nmc model (green thin line) and  for the quintessence model (black thick line)
        considered in (\cite{limagas}).  It is interesting to note
        the different behaviour of the two curves, even if the statistical significance of the best
        fit procedure for these two models is comparable: the best fit relative to our nmc model seems
        to be dominated by smaller redshift data,
        while the one relative to the Lima \& al. model by higher redshift data.     }
        \label{fgascomp}
\end{figure}

In this section we consider a recently proposed test based on the
gas mass fraction in galaxy clusters (\cite{fgasdata}). Both
theoretical arguments and numerical simulations predict that the
baryonic mass fraction in the largest relaxed galaxy clusters
should not depend on the redshift, and should provide an estimate
of the cosmological baryonic density parameter $\Omega_b$
(\cite{ENF98}). The baryonic content in galaxy clusters is
dominated by the hot X\,-\,ray emitting intra-cluster gas so that
what is actually measured is the gas mass fraction $f_{gas}$ and
it is this quantity that should not depend on the redshift.
Moreover, it is expected that the baryonic mass fraction in
clusters equals the universal ratio $\Omega_b/\Omega_M$ so that
$f_{gas}$ should indeed be given by $b {\times}
(\Omega_b/\Omega_M)$, where $\Omega_M$ is the matter density
parameter, and the multiplicative factor $b$ is motivated by
simulations that suggest that the gas fraction is lower than the
universal ratio because of processes that convert part of the gas
into stars or eject it out of the cluster altogether. Following
the procedure described in (\cite{fgasdata,allen2}), we adopt the
SCDM model (i.e., a flat universe with $\Omega_M = 1$ and $h =
0.5$, where $h$ is the Hubble constant in units of $100 \ {\rm km
\ s^{-1} \ Mpc^{-1}}$) as a reference cosmology in making the
measurements so that the theoretical expectation for the apparent
variation of $f_{gas}$ with the redshift is:
\begin{equation}
f_{gas}(z) = \frac{b \Omega_b}{(1 + 0.19 \sqrt{h}) \Omega_M}
\left [ \frac{D_A^{SCDM}(z)}{D_A^{mod}(z)} \right ]^{1.5}\,,
\label{eq: fgas}
\end{equation}
where $D_A^{SCDM}$ and $D_A^{mod}$ is the angular diameter
distance for the SCDM and our model respectively. Allen \& al.
(\cite{fgasdata}) have extensively analyzed the set of simulations
in\,(\cite{ENF98}) to get $b = 0.824 {\pm} 0.089$, so in our
analysis below, we  set $b = 0.824$. Actually, we have checked
that, for values in the $2 \sigma$ range quoted above, the main
results do not depend on $b$, as also on $\alpha_3$. Moreover we
have defined the following merit function\,:
\begin{equation} \chi^2 =
\chi_{gas}^2 + \left(\frac{\Omega_{\rm b}h^2-0.0214}{0.0020}
\right)^2 +\left(\frac{h-0.72} {0.08}
\right)^2+\left(\frac{b-0.824} {0.089} \right)^2\,, \label{eq:
defchin}
\end{equation}
where we substitute the appropriate expression of $\Omega_M$ for
our model
\begin{equation}
\chi_{gas}^2 = \sum_{i = 1}^{N_{gas}}{\left [ \frac{f_{gas}(z_i,
\alpha_1,{\widehat H}_0) - f_{gas}^{obs}(z_i)}{\sigma_{gi}} \right
]^2}. \label{eq: chigas}
\end{equation}
Here $f_{gas}^{obs}(z_i)$ is the measured gas fraction in galaxy
clusters at redshift $z_i$ with an error $\sigma_{gi}$ and the
sum is over the $N_{gas}$ clusters considered.
 Before presenting results of our analysis  it
is worth noting that, in order  to estimate the gas fraction, it is
necessary to evaluate the total cluster mass given by $M_{tot}\equiv
{M_{gas}\over f_{gas}}$. Generally the standard assumption used to
derive clusters masses from X-ray data is that the system is in
hydrostatic equilibrium. This allows one to obtain a mass estimator
just through the gas dynamical equilibrium equation:
\begin{eqnarray}
  M(<r) &=& -{r k_B T\over G \mu m_p}\left[{d\rho_{gas}\over d\ln
  r}\right]\,,\label{gaseq}
\end{eqnarray}
where $k_B$ is the Boltzmann constant, $T$ the cluster gas
temperature, $\mu$ the mean molecular weight, $m_p$ the proton
mass, and $\rho_{gas}$ the gas mass density profile. Let us note
that recently Allen \& al. (\cite{allen2}) have released a catalog
of 26 large relaxed clusters with a precise measurement of both
the gas mass fraction $f_{gas}$ and the redshift $z$. Actually to
avoid possible systematic errors in the $f_{gas}$ measurement, it
is desirable that the cluster is both highly luminous (so that the
S/N ratio is high) and relaxed, so that both merging processes and
cooling flows are absent. We use these data to perform our
likelihood analysis, getting $\chi^2=1.17$ for 26 data points, and
$\alpha_1=2.5^{+0.4}_{-0.1}$, ${\widehat H}_0=0.98\pm 0.04$,
$h=0.72\pm 0.05$, and  $w_\phi=-0.82\pm 0.1$.  To complete our
analysis we carry out a brief comparison of our results with
similar recent results of Lima et al. (\cite{limagas}), where the
equation of state characterizing the dark energy component is
constrained by using galaxy cluster x-ray data. In their analysis,
however, they consider quintessence models \textit{ in standard
gravity theories}, with a non evolving equation of state, but they
allow the so-called phantom dark energy with $w < -1$, what
violates the null energy condition. As best fit value of $w$ to
the data of (\cite{fgasdata}) they obtain
$w=-1.29_{-0.792}^{+0.686}$. In order to directly compare this
result with our analysis we first fit the model considered in
(\cite{limagas}) to the updated and wider dataset of
(\cite{allen2}), used in our analysis. To this aim we also refer
to the model function $f_{gas}(z)$, and the merit function
$\chi^2$,  defined in the Eqs. (\ref{eq: fgas}, and \ref{eq:
defchin}) respectively. We get $\chi^2=1.175$ for 26 data points,
and $\Omega_M=0.23^{+0.05}_{-0.03}$, $h=0.76^{+0.04}_{-0.09}$, and
$w=-1.11\pm 0.35$, so $w<-1$, what corresponds to a phantom
energy. We note that our model, instead, gives $w_\phi=-0.82\pm
0.1$, what does not violate the null energy condition.\, In Fig.
(\ref{fgascomp})  we compare the best fit curves for our and the
Lima \& al. model with the observational data.

%\begin{figure}
%\centering{
 %       \includegraphics[width=8 cm, height=4 cm]{fgas2.eps}}
  %      \caption{\small Best fit curve to the $f_{gas}$ data for our nmc model.}
   %     \label{fgas}
%\end{figure}
\section{Growth of density perturbations}
\begin{figure}
\centering{
        \includegraphics[width=8 cm, height=4 cm]{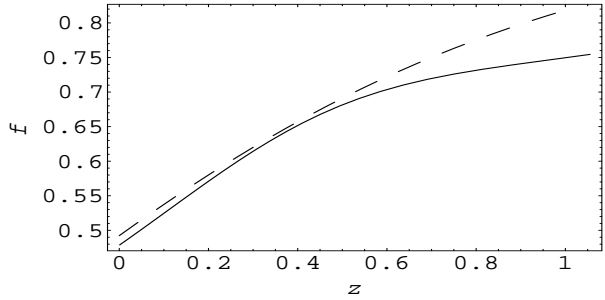}}
        \caption{\small The growth index $f$ in different cosmological models:
        the solid line corresponds to  our non minimally
        coupled model. The dashed curves correspond to a more standard quintessence
        model with an exponential potential (described in \cite{pv}).}
        \label{peculiarcompare}
\end{figure}

 In this section we consider the behavior of scalar density
 perturbations in the longitudinal gauge $ds^2= -(1 + 2
 \Phi) dt^2 + a^2 (1 - 2\Phi) d{\bf x}^2$.  Actually, while
 in the framework of the minimally coupled theory we have
 to deal with a fully relativistic component, which becomes
 homogeneous on scales smaller than the horizon, so that
 standard quintessence cannot cluster on such scales. In
 the non minimally coupled quintessence theories it is
 possible to separate a pure gravitational term both in the
 stress-energy tensor $T_{\mu \nu}$, and in the energy
 density $\rho_\phi$ , so the situation changes, and it is
 necessary to consider also fluctuations of the scalar
 field. However, it turns out (\cite{beps, uzan2}) that the
 equation for dustlike matter density perturbations inside
 the horizon can be written as follows:
\begin{equation}
{\ddot \delta_m} + 2H {\dot \delta_m} - {1\over 2} G_{\rm Cav}\,
\rho_m~\delta_m\simeq 0~,\label{del}
\end{equation}
with $G_{\rm Cav}$ is the effective gravitational constant between
two test masses and is defined by
\begin{equation}
G_{\rm Cav} = {1\over F} \left({2F+4(dF/d\phi)^2\over
2F+3(dF/d\phi)^2}\right)~. \label{Geff}
\end{equation}
The equation (\ref{del}) describes in the non minimally coupled
models evolution of the CDM density contrast, $\delta_m \equiv
\delta \rho_m /\rho_m$, for perturbations inside the horizon. In
our model the Eq.(\ref{del}) is rather complicated and takes the
form
\begin{eqnarray}\label{pertdef}
 && {\ddot \delta_m}+ \frac{2 \left(2 \alpha_2+\alpha_1 \left(-\alpha_3 e^{\alpha_1 t}-\alpha_2 t+1\right)\right)}{3
   \left(\alpha_3+e^{\alpha_1 t}+\alpha_2 t-1\right)}{\dot \delta_m}+ \\ \nonumber
   &&-\frac{e^{\alpha_1} \left(\alpha_1 \alpha_2+e^{\alpha_1} \left(\alpha_1^2-16 V_0
\right)\right)}{6 \left(\alpha_2+e^{\alpha_1}+\alpha_3-1\right)^2
   \left(\alpha_3+e^{\alpha_1 t}+\alpha_2 t-1\right) a_i^3} \delta_m=
   0,
\end{eqnarray} where
\begin{equation}\alpha_2=\displaystyle \frac{\alpha_1
\left(-\alpha_3+e^{\alpha_1}+1\right)-3 {\widehat H}_0
\left(\alpha_1+e^{\alpha_1}-1\right)}{3 {\widehat
H}_0+\alpha_1-2}.\end{equation} The Eq. (\ref{pertdef}) does not
admit exact solutions, and can be solved only numerically.
However, since with our choice of normalization the whole
history of the Universe is confined to the range $t\in[0,1]$ and
therefore to study the behavior of the solution for $t\simeq 0$
we can expand the exponential functions in Eq.~(\ref{pertdef})
in series around $t=0$. We obtain an integrable Fuchsian
differential equation, which is a hypergeometric equation. We
then use the obtained exact solution to set the initial
conditions at $t=0$ to numerically integrate  Eq.
(\ref{pertdef}) in the whole range $[0,1]$. We use the growing
mode $\delta_+$ to construct the growth index $f$ as
\begin{equation}
\lab{grow} f \equiv \frac{d \ln \delta_+ }{d \ln a}\,,
\end{equation}
where $a$ is the scale factor.
 Once we know how the growth index $f$ evolves with  redshift
and how it depends on our model parameters, we can use the
available observational data to estimate the values of such
parameters, and the present value of $\Omega_{{\rm M0}}$. The
2dFGRS team has recently collected positions and redshifts of
about $220 000$ galaxies and presented a detailed analysis of
the two-point correlation function. They measured the redshift
distortion parameter $\beta=\displaystyle{f\over b}$, where $b$
is the bias parameter describing the difference in the
distribution of galaxies and mass, and obtained that
$\beta_{|z\rightarrow 0.15}=0.49 \pm 0.09$ and $b=1.04 \pm
0.11$. From the observationally determined $\beta$ and $b$ it is
now straightforward to get the value of the growth index at
$z=0.15$ corresponding to the effective depth of the survey.
Verde \& al.~(\cite{ver+al01}) used the bispectrum of 2dFGRS
galaxies, and Lahav \& al. ~(\cite{la+al02}) combined the 2dFGRS
data with CMB data, and they obtained
\begin{eqnarray}\label{bias}
 b_{verde}&=&1.04\pm 0.11\,,\\
 b_{lahav}&=&1.19\pm 0.09\,.
\end{eqnarray}
Using these two values for $b$ we calculated the value of the
growth index $f$ at $z=0.15$, we get respectively
\begin{eqnarray}\label{peculiar}
 f_1&=&0.51\pm 0.1\,,\\
 f_2&=&0.58 \pm 0.11\,.
\end{eqnarray}
To evaluate the growth index at $z=0.15$ we first have to invert the
$z-t$ relation and find $t_{0.15}$. Then, substituting $z=0.15$ and
the two values of $f_1$ and $f_2$ we  calculate ${\widehat H}_0$ and
$\alpha_1$.  Actually the $z-t$ relation is rather involved and
cannot be exactly inverted, so we apply this procedure numerically.
We get $\alpha_1=3.3\pm 0.05$, ${\widehat
H}_0=0.98^{+0.05}_{-0.02}$,\,\,\,$V_0=0.5\pm 0.06$ which corresponds
to $\Omega_{\Lambda_{0}}=0.65\pm 0.08$. In Fig.
(\ref{peculiarcompare}) we show how the growth index is changing
with redshift in our non minimally coupled model as compared with a
standard quintessence model namely the minimally coupled exponential
model described in (\cite{pv}). We note that at low redshift
theoretical predictions of these different models are not
distinguishable, independent measurements from large redshift
surveys at different depths can disentangle this degeneracy.
\section{Conclusions}
In this paper we have shown that in the framework of non
minimally coupled scalar tensor theory of gravitation it is
possible to consider homogeneous and isotropic cosmological
models with time dependent dark energy component. These models
have a very interesting feature of producing in a natural way an
epoch of accelerated expansion. In these models initially the
over all density of the universe is dominated by matter (for the
sake of simplicity we do not include radiation into our
consideration) and later on the energy density of the scalar
field becomes dominant and the universe enters an accelerated
phase of its evolution. It turns out that at large $t$
asymptotically $w_\phi \rightarrow -1$, and such a {\it
transition} to the asymptotical value is responsible for the
accelerated expansion. The equation of state can {\it admit a
superquintessence behavior} ($w<-1$), without violating the weak
energy condition, just as an effect of the transition toward
$w\rightarrow -1$. The energy density of the scalar field
decreases with time but before the present epoch it starts to
dominate the expansion rate of the universe and asymptotically
for $t \rightarrow \infty$ the universe is reaching a de Sitter
stage.  To check viability of our model we have compared its
predictions with the available observational data and it turned
out that with an appropriate choice of parameters our model is
reproducing the observed characteristics of the universe. Let us
note that since in order to exactly solve the dynamical
equations we have not included radiation into our consideration
we do not use CMB data and observed abundances of light elements
in confrontation of theoretical predictions of our model with
observational data.

In Tab.(\ref{tab1}) we present results of our analysis, they show that
predictions of our model are fully compatible with the recent
observational data.

\begin{table*}[ht]
\begin{center}
\caption{The basic cosmological parameters derived from our model
are compared with observational data. } \label{tab1}
 \vspace{0.5cm}
 \begin{tabular}{|c|c|c|c|c|}
   \hline
$\mathbf{\alpha_1}$ & $\mathbf{{\widehat H}_0}$&$\mathbf{\Omega_{\Lambda_0}}$ & \bf{w} & \bf{dataset}\\
\hline
  $2.9^{+ 0.3}_{- 0.2}$ & $1.0^{+0.03}_{-0.04}$ & $0.84^{+0.06}_{-0.07}$& $-0.86\pm 0.06 $& Gold SNIa + SNLS\\
  \hline
   $3.2^{+0.1}_{-0.09}$ &$0.97^{+0.04}_{-0.07}$ &$ 0.62\pm 0.08 $&$-1.1\pm 0.15 $& SZe \\
   \hline
 $2.5^{+0.3}_{-0.2}$ &$ 1^{+0.05}_{-0.03}$ &$ 0.8\pm 0.15$ &$-0.76\pm 0.1$ & dimensionless coordinate\\
   \hline
     $2.8^{+0.1}_{-0.2}$&$ 1^{+0.05}_{-0.04}$ &$ 0.84\pm  0.05$&$-0.81\pm 0.07$ & GRBs\\
    \hline
   $2.5^{+0.4}_{-0.1}$ & $0.98\pm 0.04$ & $ 0.76 \pm 0.06 $  &$-0.82\pm 0.1$& gas fraction in clusters \\
   \hline
  $3.3\pm 0.05$&$ 0.98^{+0.05}_{-0.02}$ &$ 0.65\pm  0.08$&$-1.1 \pm 0.07$ & galaxies peculiar velocity\\
   \hline
    $\mathbf{3.2\pm 0.04}$&$\mathbf{ 0.99\pm +0.02}$ &$ \mathbf{0.77\pm  0.03}$&$\mathbf{-0.9\pm 0.04}$ & \bf{averaged mean}\\
    \hline
 \end{tabular}
\end{center}
\end{table*}
\section*{Acknowledgments}
This work was supported in part by the grant of Polish Ministry of
Science and Information Society Technologies 1-P03D-014-26, INFN
Na12 and the PRIN DRACO. The authors are very grateful to prof.
Djorgovski for providing the data that we used in Sec. 3.1.1. Of
course we take the full responsibility of the fitting procedure.
\section{Appendix I}
 As was already noted the solution of the coupled system of the
Einstein equations and generalized Klein-Gordon equation
describing our model can be written in the form \begin{equation}
 \label{general2}
 a(t)=A(s)\left(B(s)t^{3\over
 {s+3}}+{D\over {\Sigma_{0}}}\right)^{{s+1}\over
 s}t^{{2s^{2}+6s+3}\over {s(s+3)}}\,,
\end{equation}
\begin{equation}
 \label{generalphi2}
\phi(t)=C(s)\left(-{V_{0}\over \gamma(s)}B(s)t^{3\over
{s+3}}+{D\over \Sigma_{0}}\right)^{-{{2s+3}\over
2s}}t^{-{(2s+3)^{2}\over {2s(s+3)}}}-\phi_{0}\,,
\end{equation}
where  $A(s)$, $B(s)$, $C(s)$, $\gamma(s)$ and $\chi(s)$ are given
by
\begin{eqnarray}
  A(s)& =&\left( {\chi(s)}\right)^{s+1\over s}  \left({(s+3) \Sigma \over 3\gamma(s) }\right)^{s+2\over
  s+3}\,,
  \\
  B(s) &=& \left({(s+3) \Sigma \over 3\gamma(s) }\right)^{-{3\over (s+3)}}{(s+3)^2 \over s+6}\,,\\
  C(s) &= &\left({\chi(s)}\right)^{-{(2s+3)\over 2 s}}\left({(s+3) \Sigma \over 3\gamma(s) }\right)^{-{(3+2s)\over
  2(s+3)}},
\end{eqnarray}
and
\begin{eqnarray}
  \gamma(s) &=& {2 s+3\over 12 (s+1) (s+2)}\,,\\
  \chi(s)&=& -{  2 s\over 2 s+3}\,.
\end{eqnarray}

\end{document}